\documentclass[prl,aps,showpacs,epsf,superscriptaddress,twocolumn]{revtex4}
\usepackage{color}
\usepackage{amsmath}
\usepackage{amssymb}
\usepackage{graphicx}
\usepackage{bm}

\def\be{\begin{equation}}
\def\ee{\end{equation}}

\begin{document}

\title{Momentum distribution of a dilute unitary Bose gas with three-body losses}

\author{S\'ebastien Laurent}
\affiliation{1Laboratoire Kastler Brossel, CNRS, UPMC, Ecole Normale Sup\'erieure, 24 rue Lhomond, 75231 Paris, France}
\author{Xavier Leyronas}
\affiliation{Laboratoire de Physique Statistique, Ecole Normale Sup\'erieure, UPMC
Univ. Paris 06, Universit\'e Paris Diderot, CNRS, 24 rue Lhomond, 75005 Paris,
France.}
\author{Fr\'ed\'eric Chevy$^1$}

\begin{abstract}
Using Boltzmann's equation, we study the effect of three-body losses on the momentum distribution of a homogeneous unitary Bose gas in the dilute limit where quantum correlations are negligible. We calculate the momentum distribution of the gas and show that inelastic collisions are quantitatively as important as a second order virial correction.
\end{abstract}


\maketitle

In the past few years, ultracold gases have become a unique tool for the experimental study of strongly correlated systems. In atomic vapours, strong interactions can be achieved either by trapping the atoms in an optical lattice or by using Feshbach resonances. While the first route has been very successful and has led to ground-breaking discoveries such as the observation of the Mott transition in both Bose \cite{greiner2002quantum} and Fermi gases \cite{schneider2008metallic,jordens2008mott}, Feshbach resonances could only be used to study strongly correlated Fermi gases. Indeed, despite  interest in strongly correlated bosonic systems \cite{cowell2002cold,song2009ground,Lee2010universality,Borzov2012three,li2012bose},  the lifetime of the cloud of bosons near a Feshbach resonance is strongly reduced by the onset of three-body recombination towards deeply bound molecular states \cite{roberts2000magnetic,inouye1998observation}. Recent experimental results suggested new routes to overcome this challenge and that it might be possible to quantitatively study  the unitary Bose gas. First, it was demonstrated that at finite temperature the increase of the three-body loss rate scaling as $a^4$ actually saturates when $a\gg \lambda_{\rm th}$, where $\lambda_{\rm th}=h/\sqrt{2\pi m k_B T}$ is the thermal wavelength \cite{rem2013Lifetime,fletcher2013stability}. Moreover, recent experimental results from JILA demonstrated universal local dynamics of the momentum distribution of a unitary Bose gas  towards a quasi-equilibrium state \cite{makotyn2013universal} and have triggered several theoretical works on the dynamics of strongly correlated Bose gases near Feshbach resonances \cite{sykes2013quenching,smith2013two,yin2013quench}.

The stability of the unitary Bose gas hinges on the following argument \cite{li2012bose}: first, the three-body losses are characterized by a coefficient $L_3$ such that $\dot N=-L_3  n^2 N$, where $N$ is the total atom number and $ n$ is the particle density. This phenomenological law defines a characteristic loss rate $\gamma_3=L_3  n^2$. For a non-quantum degenerate gas, the cloud is brought back to equilibrium by elastic scattering at a characteristic rate $\gamma_2\simeq n\sigma v$, where $\sigma$ is the scattering cross-section and $v$ is the characteristic  velocity of the atoms. At unitarity, the scattering cross-section follows a universal scaling $\sigma=8\pi/k^2$, where $k$ is the relative wave-vector of two scattering particles. In the presence of losses, the system can be kept in a quasi-equilibrium state provided that the ratio $\gamma_3/\gamma_2$ stays small. It was shown both theoretically and experimentally \cite{rem2013Lifetime,fletcher2013stability} that at unitarity the three body loss-rate is given by
\be
L_3\simeq 36\sqrt{3}\pi^2\frac{\hbar^5}{m^3(k_B T)^2}(1-e^{-4\eta}),
\label{Eq0}
\ee
where $\eta$ is a dimensionless parameter characterizing the probability of forming a deeply bound molecule at short distance \cite{braaten2003Universal}. Plugging Eq. (\ref{Eq0}) into the expression for $\gamma_3$, we see that quasi-equilibrium can be achieved as long as $(1-e^{-4\eta})n\lambda_{\rm th}^3$ is small, i.e. when the system is not too deeply in the quantum degenerate regime.

In this letter, we investigate the effect of 3-body losses on the momentum distribution of a unitary Bose gas. Our analysis is based on a semi-analytical resolution of Boltzmann's equation. Since Boltzmann's equation neglects all many-body correlations, our work is restricted to a low-phase space density regime where, as aforementioned, three-body losses can be treated perturbatively.  We calculate the first correction to the momentum distribution and we compare it to the effect of two-body interactions. We show that in the dilute limit, both effects deplete the center of the momentum distribution proportionally to the phase-space density of the gas. Moreover, for realistic parameters, this depletion is dominated by three-body losses.

Consider a homogeneous Bose gas that we describe by a phase space density $f(\bm p)$. In the presence of losses, $f$ is the solution of Boltzmann's equation that we write formally
\be
\partial_t f=I_{\rm coll}[f]-{\cal L}_3[f],
\label{Eq1}
\ee
where $I_{\rm coll}$ and ${\cal L}_3$ are non linear operators  describing respectively the elastic collisions and the three-body losses. At low phase space density, we can neglect the bosonic stimulation and we have
\be
I_{\rm coll}[f](\bm p_1)=\int {\rm d}^3\bm p_2 {\rm d}^2\bm\omega' \frac{{\rm d}\sigma}{{\rm d}\omega'}\frac{|\bm p_2-\bm p_1|}{m}\left(f_3f_4-f_1f_2\right).
\ee
Here, $f_\alpha$ stands for $f(\bm p_\alpha)$, $(\bm p_1,\bm p_2)$ (resp. $(\bm p_3,\bm p_4)$) are the incoming (outgoing) momenta satisfying energy and momentum conservation and ${\rm d}\sigma/{\rm d}\omega'=8\hbar^2/|\bm p_1-\bm  p_2|^2$ is the differential scattering cross-section towards the outgoing solid angle $\bm \omega'$.

From \cite{rem2013Lifetime}, the loss rate operator for a unitary Bose gas can be written as
\be
{\cal L}_3[f](\bm p_1)=\int {\rm d}^3\bm p_2{\rm d}^3\bm p_3\frac{{\cal A}_3}{E_{123}^2}|\phi(\bm \Omega_3)|^2f(\bm p_1)f(\bm p_2)f(\bm p_3),
\ee
where $E_{123}=(p_1^2+p_2^2+p_3^2)/2m-(\bm p_1+\bm p_2+\bm p_3)^2/6m$ is the energy in the center of mass frame of the three particles of momenta $(\bm p_1,\bm p_2,\bm p_3)$, ${\cal A}_3=2\pi^3(k_{\rm B}T)^2L_3$ and $\phi(\bm \Omega_3)$ is the hyperangular  wave-function describing the angular structure of the Efimov trimers that we normalize by the condition $\int {\rm d}^5\bm\Omega_3|\phi(\bm\Omega_3)|^2=1$.

In absence of losses the system thermalizes to a distribution $G$ solution of $I_{\rm col}[G]=0$. For a classical gas, the solution of this equation is a Gaussian distribution $G(n,E; p)=n\lambda^3_{\rm th}e^{-\beta p^2/2m}/h^3$, where $\beta =1/k_B T$ and $E=\int  (G(p) p^2/2m) d^3\bm p=3nk_B T/2$ is the energy density.

In the quasi-static regime $\gamma_3/\gamma_2\ll 1$, three-body losses are small and we can use ${\cal A}_3$ as an expansion parameter. Since for ${\cal A}_3=0$ the system can reach a stationary thermal state, we expect the characteristic evolution time in the presence of losses to vary as ${\cal A}_3^{-1}$ and thus $\partial_t$ must be considered to scale as ${\cal A}_3$.
We write  then $f=f_0+f_1+...$ where $f_j\propto {\cal A}_3^j$. The expansion of Eq. (\ref{Eq1}) to first order in ${\cal A}_3$, yields
\begin{eqnarray}
I_{\rm coll}[f_0]&=&0\label{Eq2}\\
\partial_t f_0&=&I'_{\rm coll}[f_1]-{\cal L}_3[f_0].\label{Eq3}
\end{eqnarray}
where $I'_{\rm coll}$ is the linearized collisional operator.

According to (\ref{Eq2}), $f_0$ is a Maxwell-Boltzmann distribution. However, since the system loses particles by three-body recombination, its atom number and its energy vary with time. We therefore have $f_0(p,t)=G(n_t,E_t;p)$. We then have in Eq. (\ref{Eq3})
\be
I'_{\rm coll}[f_1]={\cal L}_3[f_0]+\dot E\partial_E G+\dot n \partial_n G.
\label{Eq5}
\ee

Take $f_1(p,t)=G(n_t,E_t;p) \alpha(p,t)$. Eq. (\ref{Eq5}) then becomes
\be
C[\alpha]=\frac{1}{G}{\cal L}_3[G]+\dot E\partial_E \ln(G)+\dot n \partial_n \ln(G).
\label{Eq6}
\ee
with
\begin{eqnarray}
C[\alpha]&=&\frac{1}{G}I'_{\rm coll}[G \alpha]\\
&=&\int {\rm d}^3\bm p_2{\rm d}^2\bm\omega' f_0(\bm p_2)\frac{{\rm d}\sigma}{{\rm d}\omega'}\frac{|\bm p_2-\bm p_1|}{m}\nonumber\\
&&\times\left(\alpha_3+\alpha_4-\alpha_1-\alpha_2\right).
\end{eqnarray}
and $\alpha_k=\alpha(\bm p_k)$ for $k=1,\cdots 4$. The operator $C$ is symmetric for the scalar product \cite{smith1989transport}
\be
\langle \alpha|\alpha'\rangle=\int {\rm d}^3\bm p G(p)\alpha(p)\alpha'(p).
\label{Eq4}
\ee
Due to energy and particle number conservation, the kernel of $C$ is spanned by $\alpha(p)=1$ and $\alpha(p)=p^2$. Finally, being a symmetric operator, its image is orthogonal to its kernel.
To find the time evolution of the energy and the atom number, we project Eq. (\ref{Eq6}) on $1$ and $p^2$. Using the structure of the kernel of $C$, the collisional term vanishes and  we obtain
\begin{eqnarray}
\dot n_t&=&-\langle 1|\frac{1}{G}{\cal L}_3[G]\rangle\\
\dot E_t&=&-\langle \frac{p^2}{2m}|\frac{1}{G}{\cal L}_3[G]\rangle.
\end{eqnarray}
The explicit calculation of the rhs of these equations involves 9-dimensional integrals over the three momenta $(\bm p_1,\bm p_2,\bm p_3)$ in the three-body loss rate operator. This calculation can be performed analytically by introducing the momentum-space Jacobi coordinates (see supplemental material) and we finally obtain
\begin{eqnarray}
\dot n_t&=&-L_3n^3
\label{Eq621}\\
\dot E_t&=&-\frac{5}{9} E L_3 n^2.
\label{Eq622}
\end{eqnarray}
where we recover the usual formula for three-body losses, as well as the recombination heating discussed in \cite{rem2013Lifetime,fletcher2013stability}.

To find $\alpha$, we project Eq. (\ref{Eq6}) on the range of $C$ (ie orthogonally to ${\rm Span}(1,p^2)$). We then have
\be
C[\alpha]=P\left[\frac{1}{G}{\cal L}_3[G]\right],
\label{Eq7}
\ee
where $P$ is the orthogonal projector on ${\rm Im}(C)$, and where we used the fact that $\ln G$ is a linear combination of $1$ and $p^2$ and thus lies in the kernel of $C$ and $P$.

\begin{figure}
\centerline{\includegraphics[width=\columnwidth]{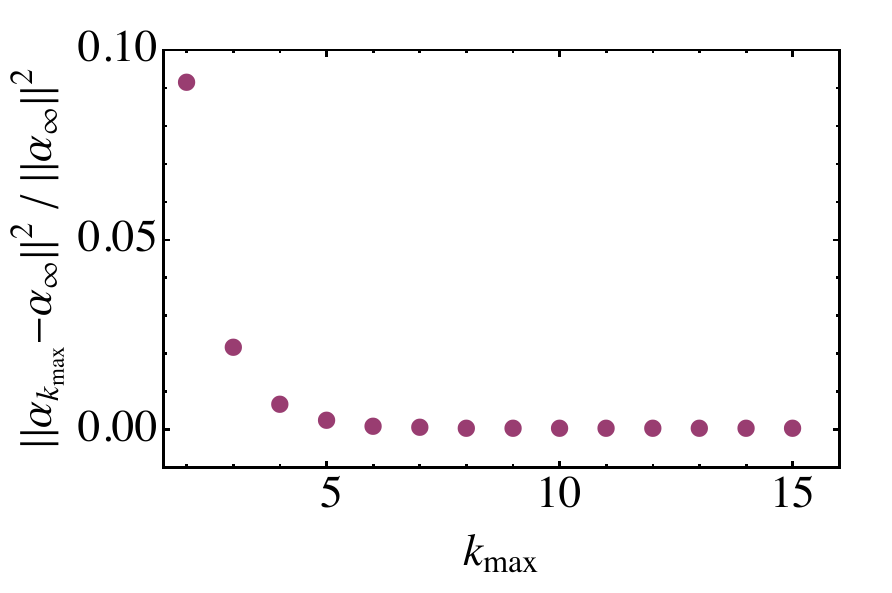}}
\caption{Convergence of the numerical solution of Eq. (\ref{Eq8}). We estimate the error on the solution using the norme $\|\alpha\|^2=\langle\alpha|\alpha\rangle$ and we compare the solution of Eq. (\ref{Eq8}) obtained by truncation at $k=k_{\rm max}$ with the ``true" result corresponding to $k_{\rm max}=15$. }
\label{Fig1}
\end{figure}

In the spirit of Chapman-Enskog's expansion, we expand $\alpha$ on a basis of orthogonal polynomials for the scalar product (\ref{Eq4}). Such a basis can be expressed in terms of the generalized Laguerre polynomials \cite{abramowitz1970handbook}
\be
q_k(p)=\sqrt{\frac{\sqrt{\pi}k!}{2n\Gamma(k+3/2)}}L_k^{(1/2)}(\beta p^2/2m)
\ee
By definition, $q_0$ and $q_1$ lie in ${\rm Ker}(C)$ and as such will not contribute to the expansion. Take $\alpha(p)=\sum_{k\ge 2}a_k q_k(p)$, where the coefficients $a_k$ are real numbers, Eq. (\ref{Eq7}) is then equivalent to  the infinite set of linear equations
\be
\sum_{k'\ge 2}a_{k'}\langle q_k|C[q_{k'}]\rangle=\langle q_k|\frac{1}{G}{\cal L}_3[G]\rangle,
\label{Eq8}
\ee
for $k\ge 2$. In these equations, the coefficients $\langle q_k|C[q_{k'}]\rangle$ can be calculated analytically to arbitrary order (see supplemental material), while the complex form of the Efimov wave-function allows only for a numerical calculation of the projection of the loss term on this polynomial basis.  We solve this equation by truncating the indices $(k,k')$ to a value $k_{\rm max}$. We observe in Fig. (\ref{Fig1}) that the convergence is very fast and that the first order result ($k_{\rm max}=2$) gives the correct answer within a few percent accuracy.

\begin{figure}
\centerline{\includegraphics[width=\columnwidth]{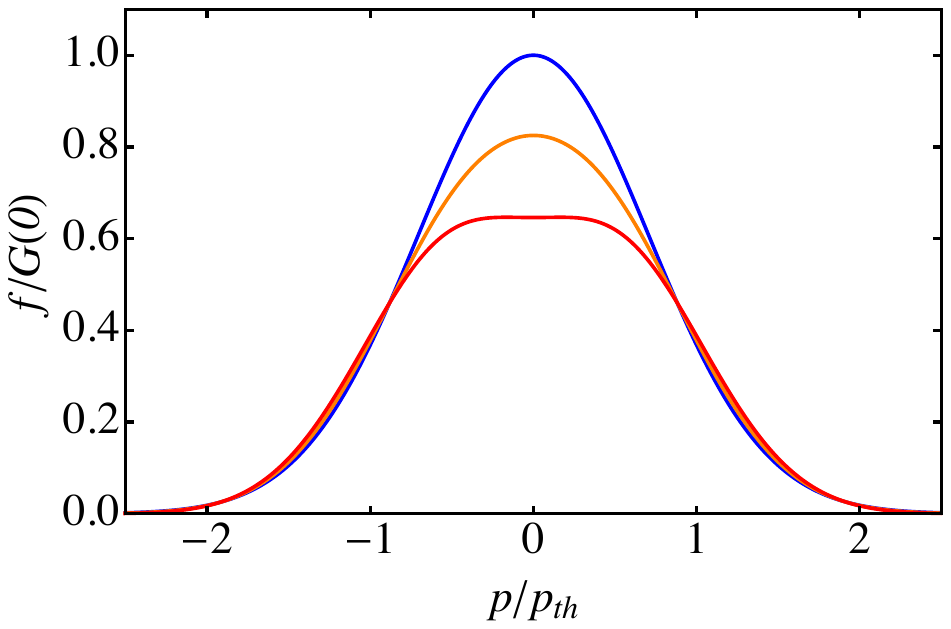}}
\caption{Color online. Deformation of the momentum distribution of a unitary Bose gas due to three-body losses. From top to bottom: $n\lambda_{\rm th}^3(1-e^{-4\eta})=0$ (Blue, Boltzmann gas); $n\lambda_{\rm th}^3(1-e^{-4\eta})=0.05$ (Orange) and $n\lambda_{\rm th}^3(1-e^{-4\eta})=0.1$ (Red).}
\label{Fig2}
\end{figure}

In an experiment such as the one described in \cite{makotyn2013universal}, the cloud is not directly prepared in the quasi-static, strongly interacting state. Rather, the experimental sequence starts in the weakly interacting regime where losses can be neglected and the momentum distribution of the gas is gaussian. The magnetic field is then ramped quickly to unitarity where the system can relax towards the quasi-equilibrium described above. To study the relaxation rate towards the quasi-static solution we write as before $f=f_0+f_1$ with $f_1=f_{1,{\rm qs}}+\delta f_1$, where $f_{1,{\rm qs}}$ is the quasi-static solution and $\delta f_1$satisfies the initial condition $\delta f_1(p,t=0)=-f_{1,{\rm qs}}(p;t=0)$, since at $t=0$, $f=f_0$. Expanding Boltzmann's Equation to first order in $f_1$ and using the properties of $f_{1,{\rm qs}}$, we obtain for $\delta f_1$

\be
\partial_t \delta f_1=I'_{\rm coll}[\delta f_1].
\ee
This equation shows that the relaxation towards the quasi-static regime is solely driven by two-body collisions and occurs at a rate $\sim\gamma_2$. This may seem paradoxical since one would rather expect the three-body characteristic rate $\sim\gamma_3$ . However, as far as the phase-space density is concerned, the depletion of $f$ at low momenta is quite small since the relative decrease of the peak momentum density is $\propto n\lambda^3$. Since $1/\gamma_3$ is the time required to lose typically half the initial atom number, the dip should form on a time scale $\simeq n\lambda^3/\gamma_3\simeq 1/\gamma_2$.

The three-body losses lead to a correction to the momentum distribution proportional to $n\lambda^3$. This scaling is similar to the first virial correction, and one may wonder if the three-body losses might not mask the effects of two-body interactions. To clarify this point, we calculated the leading order corrections to the occupation number $\rho(p)=h^3 f(p)$ using the scheme presented in \cite{leyronas2011virial}. In the virial expansion, the leading order term corresponds to the ideal Boltzmann gas. In the grand canonical ensemble, this term reads $\rho^{(1)}(p)=ze^{-\beta\varepsilon_p}$, where $z$ is the fugacity and  $\varepsilon_p=p^2/2m$. The next order term is the sum of two contributions. The first one corresponds to Bose's statistics and is simply $\rho^{(2,a)}(p)=z^2 e^{-2\beta \varepsilon_p}$ while the second one is more involved and is due to interactions. Following \cite{leyronas2011virial}, it is given by
\be
\begin{split}
\rho^{(2,b)}(p)=\frac{8\pi z^2}{m}
\int_{\mathcal C_{\gamma}}\frac{{\rm d}s}{2\pi i}\int_{0}^{+\infty}\frac{{\rm d}P P^2}{2\pi^2} \frac{e^{-\beta s}}{\sqrt{-m s}}\\
\times
\frac{e^{-\beta \frac{P^2}{4m}}}
{
\left[s+\frac{P^2}{4m}-\frac{p^2}{2m}-\frac{(P-p)^2}{2m}
\right]
\left[s+\frac{P^2}{4m}-\frac{p^2}{2m}-\frac{(P+p)^2}{2m}
\right]
}
\end{split}
\label{Eq9}
\ee
where ${\mathcal C_{\gamma}}$ is a Bromwich contour \cite{appel2007mathematics}. We note that this expression is simply twice that obtained for spin 1/2 fermions \cite{leyronas2011virial}. To convert this momentum distribution to the canonical ensemble, we use the virial expansion of the equation of state of the unitary Bose gas, $n\lambda_{\rm th}^3=z+2 b_2z^2+...$, with $b_2=9/4\sqrt{2}$. We thus obtain

\be
\rho(p)=n\lambda_{\rm th}^3e^{-\beta\varepsilon_p}+(n\lambda_{\rm th}^3)^2\left[\xi(\lambda_{\rm th} p/\hbar)-2 b_2 e^{-\beta\varepsilon_p}\right],\label{eqvir}
\ee
where we took $\rho^{(2)}(p)=\rho^{(2,a)}(p)+\rho^{(2,b)}(p)=z^2\xi(\lambda_{\rm th} p/\hbar)$.

\begin{figure}
\centerline{\includegraphics[width=\columnwidth]{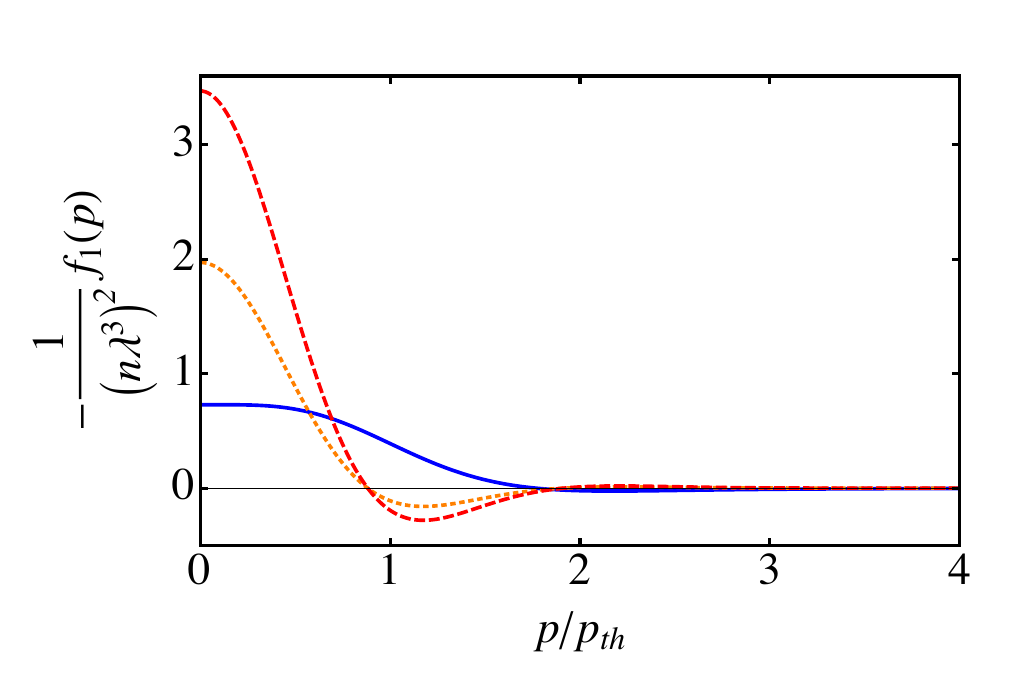}}
\caption{Correction to the Boltzmann gas: three-body losses vs interactions. The correction to Boltzmann's distribution is plotted for maximal three-body losses ($\eta=\infty$, red dashed line), $\eta=0.2$, corresponding to $^7$Li (Orange dotted line). The blue solid line corresponds to the correction Eq. (\ref{eqvir}) due to Bose statistics and two-body interactions.}
\label{Fig3}
\end{figure}

In Fig. \ref{Fig3}, we compare the effect of 3-body losses with the virial corrections to the momentum distribution. We observe that for $^7$Li, for which $\eta=0.2$, the dip in the momentum distribution is dominated by three-body losses.

From a quantitative point of view, the analysis presented above gives controlled results in the high temperature regime since at the lowest order in phase space density,  we expect the two corrections ($3$-body losses and virial expansion) to be additive. From a more qualitative point of view, since at large temperature the ratio between $3$-body losses and two-body collision rates is small, one may naively assume that  the effect of the losses on the momentum distribution function to be superseded by quantum correlations effects (interactions and bosonic statistics). Surprisingly, we find on the contrary that they scale identically with $n\lambda_{\rm th}^3$ and that for typical values of the parameter $\eta$, the losses actually dominate.

If we decrease the temperature, the ratio $\gamma_3/\gamma_2$ increases, and as a consequence we anticipate a more important role of $3$-body losses. Therefore we think the phenomenon we discuss in the present work will become more important as the temperature decreases. As such, it may have an important role in the interpretation of the results presented in \cite{makotyn2013universal}. One may argue that the experiments presented in this reference were obtained after a time short compared to the three-body lifetime. However, as  shown above, even in the limit $\gamma_2\gg \gamma_3$, the relaxation towards the quasi-static distribution driven by three-body recombination occurs on a time-scale $1/\gamma_2\ll 1/\gamma_3$.

Finally, even though our work is restricted to a homogeneous system while real experiments are usually performed in harmonic traps, we note that if the trapping frequency is small enough (as in the experiment of \cite{makotyn2013universal}), the elastic and inelastic relaxation times can become shorter than the typical atomic diffusion time. In this case, both the elastic and inelastic dynamics occur locally, and can therefore be described using our formalism.

\acknowledgments

The authors thank A. Grier, C. Salomon, F. Werner and the ENS cold atom group for careful reading and helpful discussions. The authors acknowledge support from the ERC (Advanced grant Ferlodim and starting grant Thermodynamix), Institut Universitaire de France and R\'egion Ile de France (IFRAF). FC thanks Laure Saint-Raymond for fruitful discussions.

\bibliographystyle{unsrt}
\bibliography{bibliographie}

\newpage

\section{Derivation of the loss equations}

The coefficient $\langle 1|\frac{1}{G}{\cal L}_3[G]\rangle$ can be written as

\be
\begin{split}
\langle 1|\frac{1}{G}&{\cal L}_3[G]\rangle=\\&
\left(\frac{n\lambda^3_{\rm th}}{h^3}\right)^3\int {\rm d}^3\bm p_1{\rm d}^3\bm p_2{\rm d}^3\bm p_3\frac{{\cal A}_3}{E_{123}^2}|\phi(\bm \Omega_3)|^2e^{-\beta E_{\rm tot}}
\end{split}
\ee
where $E_{\rm tot}=(p_1^2+p_2^2+p_3^2)/2m$.
We then define three new momentum variables which are conjugated to Jacobi coordinates in real space and verify
\begin{eqnarray}
\bm p_1&=& \frac{\bm P}{3}-\frac{\bm \Pi_1}{a}-\frac{a\bm \Pi_2}{2}\\
\bm p_2&=& \frac{\bm P}{3}+\frac{\bm \Pi_1}{a}-\frac{a\bm \Pi_2}{2}\\
\bm p_3&=& \frac{\bm P}{3}+a\bm \Pi_2.
\end{eqnarray}
with $a =(4/3)^{1/4}$.\\
The energy in the center of mass frame is then $E_{123}=\Pi^2/2\mu$ with $\Pi^2=\Pi_1^2+\Pi_2^2$ and $\mu=m/\sqrt{3}$ while the total energy is $E_{\rm tot}=P_{\rm G}^2/6m+\Pi^2/2\mu$.
The jacobian of such a change of variables is equal to one and we have the differential transformation
\be
{\rm d}^3\bm p_1{\rm d}^3\bm p_2{\rm d}^3\bm p_3={\rm d}^3\bm P_{\rm G}\Pi^5{\rm d} \Pi\frac{1}{2}{\rm sin}^2(2\alpha) {\rm d} \alpha {\rm d}^2\hat{\Pi}_1 {\rm d}^2\bm \hat{\Pi}_2
\ee
where $\hat{\Pi}_{i}=\bm \Pi_{i}/\Pi_{i}$ and $\alpha=\mathrm{arctan}(\Pi_{1}/\Pi_{2})\in [0 ;\pi/2 ]$.\\ It can be rewritten in terms of the hyperangular differential $\mathrm{d}^5\bm \Omega_3=1/2\mathrm{sin}^2(\alpha)\mathrm{d}\alpha\mathrm{d}^2\hat{\Pi}_{1}\mathrm{d}^2\hat{\Pi}_{2}$.\\
We thus obtain a new form for the integral
\be
\begin{split}
\langle &1|\frac{1}{G}{\cal L}_3[G]\rangle=\\&\left(\frac{n\lambda^3_{\rm th}}{h^3}\right)^3\int {\rm d}^3\bm P_{\rm G}\Pi^5{\rm d} \Pi \mathrm{d}^5\bm \Omega_3 \frac{{\cal A}_3}{E_{123}^2}|\phi(\bm \Omega_3)|^2e^{-\beta E_{\rm tot}}.
\end{split}
\ee
Using the normalization condition on $\phi(\bm \Omega_3)$ we are left with Gaussian integrals which are straightforward to calculate. We then recover easily (\ref{Eq621}).\\
To calculate $\langle \frac{p^2}{2m}|\frac{1}{G}{\cal L}_3[G]\rangle$  we use the fact that it can be written as

\be
\begin{split}
\langle &\frac{p_1^2}{2m}|\frac{1}{G}{\cal L}_3[G]\rangle=\\&
\left(\frac{n\lambda^3_{\rm th}}{h^3}\right)^3\int {\rm d}^3\bm p_1{\rm d}^3\bm p_2{\rm d}^3\bm p_3\frac{E_{\rm tot}}{3}\frac{{\cal A}_3}{E_{123}^2}|\phi(\bm \Omega_3)|^2e^{-\beta E_{\rm tot}}.
\end{split}
\ee
 Therefore we can use the same change of variables to get rid of the hyperangular dependence and finally retrieve  (\ref{Eq622}).

\section{Calculation of $C$}

The coefficients $\langle q_k|C[q_{k'}]\rangle$ can be expressed as follow $\langle q_k|C[q_{k'}]\rangle=-n \hbar ^2\sqrt{\pi \beta /m^{3}} A_{kk'} $, $A=(A_{kk'})$ being a matrix with purely numerical coefficients. Those coefficients can be calculated analytically to arbitrary order. As a ``proof", all the coefficients to a value $k_{max}=6$ are shown below:
\be
A=\left(
\begin{array}{cccccc}
 0 & 0 & 0 & 0 & 0 & 0 \\
 0 & 0 & 0 & 0 & 0 & 0 \\
 0 & 0 & \frac{256}{45} & \frac{64}{15} \sqrt{\frac{2}{21}} & \frac{32}{15 \sqrt{21}} &
   \frac{8}{9} \sqrt{\frac{10}{231}} \\
 0 & 0 & \frac{64}{15} \sqrt{\frac{2}{21}} & \frac{288}{35} & \frac{8 \sqrt{2}}{5} &
   \frac{428}{63 \sqrt{55}} \\
 0 & 0 & \frac{32}{15 \sqrt{21}} & \frac{8 \sqrt{2}}{5} & \frac{14908}{1575} & \frac{533}{35}
   \sqrt{\frac{2}{55}} \\
 0 & 0 &\frac{8}{9} \sqrt{\frac{10}{231}}  & \frac{428}{63 \sqrt{55}} & \frac{533}{35}
   \sqrt{\frac{2}{55}} & \frac{209863}{20790}
  \end{array}
\right)
\\
\ee

\end{document}